\documentstyle[eqsecnum,aps,manuscript]{revtex}

\def\b{\begin{equation}}
\def\e{\end{equation}}
\begin{document}
\draft

\title{VANISHING CORRECTIONS ON INTERMEDIATE SCALE AND IMPLICATIONS 
FOR UNIFICATION  OF FORCES}

\author{M.K.Parida}

\address{High Energy Physics Group\\ 
International Center for Theoretical Physics\\
I-34100 Trieste,Italy\\
and\\
Physics Department\\
North--Eastern Hill University\\
P.O.Box 21, Laitumkhrah, Shillong 793 003\\
India}

\date{March 19, 1996}

\maketitle

\begin{abstract}
In two--step breaking of a class of grand unified theories including
$SO(10)$, we prove a theorem showing that the scale $(M_I)$ where the 
Pati--Salam gauge symmetry with parity breaks down to the standard gauge 
group, has vanishing corrections due to all sources emerging from higher 
scales $(\mu>M_I)$ such as the one--loop and all higher--loop effects, 
the GUT--threshold, gravitational smearing, and string threshold effects. 
Implications of such a scale for the unification of gauge couplings 
with small Majorana neutrino masses are discussed. In string inspired 
$SO(10)$, we show that $M_I\simeq 5\times 10^{12}$ GeV, 
needed for neutrino masses, with the GUT scale 
$M_U\simeq M_{str}$ can be realized provided certain particle states 
in the predicted spectrum are light.
\end{abstract}
\pacs{12.10.Dm, 12.60.Jv}

\narrowtext

\section{INTRODUCTION}

Grand unified theories based upon SUSY $SU(5),SO(10)$, nonSUSY $SO(10)$ 
with intermediate symmetries, and those inspired by superstrings have 
been the subject of considerable interest over recent years. 
In order to solve the strong CP problem through Peccei--Quinn mechanism 
and achieve small neutrino masses [1] necessary to understand the solar 
neutrino flux [2] and/or the dark matter of the universe, an intermediate 
scale seems to be essential [3]. Such a
scale might correspond to the spontaneous breaking of gauged 
B--L contained in intermediate gauge symmetries like 
$SU(2)_L\times SU(2)_R\times U(1)_{B-L}\times SU(3)_C(\equiv G_{2213})$ 
and $SU(2)_L\times SU(2)_R\times SU(4)_C\,(\equiv G_{224})$ with
[3-6] or without [7] parity, or even others like 
$SU(2)_L\times U(1)_{I_{3R}}\times SU(4)_C$ and
$SU(2)_L\times U(1)_{I_{3R}}\times U(1)_{B-L}\times SU(3)_C$. 
But it is well known that the predictions of a grand unified theory 
are more [8] or less [6,9] 
uncertain predominantly due to threshold [10] and gravitational 
smearing effects [11,12] originating from higher dimensional operators. 
The uncertainty in the intermediate scale prediction naturally leads 
to theoretical uncertainties in the neutrino mass 
predictions through seesaw mechanism. Therefore, an intermediate scale, 
stable against theoretical uncertainties, would be most welcome from 
the point of more accurate predictions on neutrino masses.
\par 
Another problem in SUSY GUTs having supergrand desert is 
the requirement of $\alpha_s(M_Z)\geq 0.12$ to achieve unification at 
$M_U\simeq 2\times 10^{16}$ GeV. Even though the problem is 
alleviated by unknown GUT threshold and gravitational corrections [13], 
realization of a natural grand unification scale 
$M_U\simeq M_{str}\simeq 5.6\times g_{str}\times 10^{17}$ GeV 
requires the presence of some lighter string states which could 
be the extra gauge bosons or Higgs scalars of a unifying symmetry, 
exotic vector--like quarks and leptons with nonconventional 
hypercharge assignments [14-16], or a $SU(3)_C$--octet and weak 
$SU(2)$--triplet in the adjoint representation of the standard 
gauge group [17]. But, in the absence of an intermediate symmetry, 
the neutrino mass predictions may fall short of the solar 
flux requirements by 2-3 orders. Assuming boundary conditions at the 
string scale to be different from a GUT--boundary condition, attempts 
have been made to bring down the values of intermediate scales relevant 
for larger neutrino masses [18].
\par 
The presence of a $G_{224P}$ intermediate gauge symmetry, having 
only two couplings for $\mu >M_I$, would always guarantee gauge 
unification, and a demonstration of $M_I\simeq 10^{12}-10^{14}$ GeV 
with $M_U\simeq M_{str}$ in SUSY inspired $SO(10)$, would 
solve at least two of the major problems: the string scale unification 
with $\alpha_s(M_Z)\simeq 0.11$ and neutrino masses needed for solar 
neutrino flux.
\par 
It has been shown recently that in all GUTs where $G_{224P}$ breaks 
spontaneously at the highest intermediate scale, the 
$\sin^2\theta_W(M_Z)$ prediction is unaffected by GUT--threshold 
and multiloop (two--loop and higher) radiative corrections emerging 
from higher mass scales [6]. As a single intermediate symmetry 
is more desirable from minimality consideration, we confine to the
single $G_{224P}$ symmetry in 
two--step breakings of all possible GUTs including $SO(10)$ 
and prove a theorem showing that all higher--scale corrections to 
the intermediate scale $(M_I)$ prediction vanish. In SUSY $SO(10)$ 
inspired by superstrings [19], we find that $M_I\simeq 10^{12}-10^{14}$ 
GeV is possible with $M_U\simeq M_{str}$ provided certain states in 
the predicted spectrum are light.

\section{THEOREM ON VANISHING CORRECTIONS ON THE INTERMEDIATE SCALE}

We now state the following theorem and provide its proof,\\ 
{\bf Theorem:} {\it In all two--step breakings of grand unified theories, 
the mass scale $(M_I)$ corresponding to the spontaneous breaking of the 
intermediate gauge symmetry 
$SU(2)_L\times SU(2)_R\times SU(4)_C\times P(g_{2L}=g_{2R})$,
has vanishing contributions due to every correction term 
emerging from higher scales $(\mu>M_I)$.}
\par 
To prove the theorem we consider the two--loop breaking pattern in SUSY 
or nonSUSY GUTs,
\[{\rm GUT}\buildrel M_U\over\longrightarrow G_{224P}\buildrel M_I\over
\longrightarrow G_{213}\buildrel M_Z\over
\longrightarrow U(1)_{em}\times SU(3)_C\nonumber\] 
which may or may not originate from superstrings. 
Following the standard notations, we use the following renormalization 
group equations (RGEs) for the gauge couplings 
$\alpha_i(\mu)=g_i^2(\mu)/4\pi$,\\
$\underline{M_Z\leq\mu\leq M_I}$
\FL\[{1\over\alpha_i(M_Z)}=
{1\over\alpha_i(M_I)}+{a_i\over 2\pi}\ln{M_I\over M_Z}+
\theta_i-\triangle_i\;,\nonumber\]
\FR\b i=Y,2L,3C\e
$\underline{M_I\leq\mu\leq M_U}$
\FL\[{1\over\alpha_i(M_I)}={1\over\alpha_i(M_U)}+{a'_i\over 2\pi}
\ln{M_U\over M_I}+
\theta'_i-\triangle'_i\;,\]
\FR\b i=2L,2R,4C\e 
where $\triangle_i$ includes threshold effects at 
$\mu=M_Z(\triangle^Z_i)$ due to the top--quark and Yukawa 
couplings and superpartners in SUSY theories. It also includes 
threshold effects $(\triangle^I_i)$ due to
heavy particles near the intermediate scale,
\b\triangle_i=\triangle_i^{(Z)}+\triangle_i^{(I)}\;,\;i=Y,2L,3C\e 
The second (third) term in the r.h.s. of (2.1)--(2.2) is the usual 
one--loop (multiloop) contribution.
\par 
The GUT threshold $(\triangle_i^U)$,  
gravitational corrections $(\triangle_i^{NRO})$, or the string 
threshold effects $(\triangle_i^{str})$ when the model is based upon 
string inspired $SO(10)$ [20], are contained in $\triangle'_i$,
\b\triangle'_i=\triangle_i^U+\triangle_i^{NRO}+\triangle_i^
{str}\;,\;i=2L,2R,4C\e 
In nonSUSY and SUSY GUTs, the $\triangle_i^{NRO}$ may emerge from 
higher dimensional operators scaled by the Planck mass [11] leading 
to a nonrenormalizable Lagrangian
\FL\[{\cal L}_{NRO}=\nonumber\]
\b-{\eta^{(1)}\over 2M_{pl}}{\rm Tr}
(F_{\mu\nu}\phi F^{\mu\nu})-{\eta^{(2)}\over 2M^2_{pl}}{\rm Tr}
(F_{\mu\nu}\phi^2 F^{\mu\nu})+\ldots\e 
where $M_{Pl}=$ Planck mass, and $\phi=$ Higgs field which is 
responsible for breaking the GUT symmetry to $G_{224P}$. For example, 
in $SO(10),\,\phi=\underline{54}$. These operators lead to the 
modifications of the GUT--scale boundary conditions on gauge couplings,
\FL\[\alpha_{2L}(M_U)(1+\in_{2L})=\alpha_{2R}(M_U)(1+\in_{2R})
=\nonumber\]
\FR\b\alpha_{4C}(M_U)(1+\in_{4C})=\alpha_G\e 
which imply\b\triangle_i^{NRO}=-{\in_i\over\alpha_G}\;,\;i=2L,2R,4C\e 
where $\alpha_G=$ GUT coupling and $\in_i$ are known functions of 
the parameters $\eta^{(i)}$, the vacuum expectation value of 
$\phi$, $M_U$, and $M_{Pl}$.
\par 
Using suitable combinations of gauge 
couplings and eqs.(2.1)--(2.2), we obtain the following analytic 
formulas,
\FL\[\ln{M_U\over M_Z}={(L_SB_I-L_\theta A_I)\over D}+\nonumber\]\FR
\b{(J_\theta B_I-K_\theta A_I)\over D}+{(K_\triangle
A_I-J_\triangle B_I)\over D}\e
\FL\[\ln{M_I\over M_Z}={(L_\theta A_U
-L_SB_U)\over D}+\nonumber\]
\FR\b{(K_\theta A_U-J_\theta B_U)\over D}+{(J_\triangle B_U-
K_\triangle A_U)\over D}\e\[ D=A_UB_I-A_IB_U\nonumber\]\[ L_S=
{16\pi\over 3\alpha (M_Z)}\left[\;{\alpha(M_Z)\over \alpha_S(M_Z)}-
{3\over 8}\;\right]\nonumber\]
\b L_\theta={16\pi\over 3\alpha(M_Z)}\left[\;\sin^2\theta_W(M_Z)-
{3\over 8}\;\right]\e\[ A_U=2a'_{4C}-a'_{2L}-a'_{2R}\nonumber\]
\[ B_U={5\over 3}a'_{2L}-a'_{2R}-{2\over 3}a'_{4C}\nonumber\]
\[ A_I={8\over 3}a_{3C}-a_{2L}-{5\over 3}a_Y-A_U\nonumber\]
\[ B_I={5\over 3}(a_{2L}-a_Y)-B_U\nonumber\]
\[ J_\theta=2\pi\left[\;\theta_{2L}+{5\over 3}\theta_Y-{8\over 3}
\theta_{3C}+\theta'_{2L}+\theta'_{2R}-2\theta'_{4C}\;\right]\nonumber\]
\[ K_\theta =2\pi\left[\;{5\over 3}(\theta_Y-\theta_{2L})+
\theta'_{2R}+{2\over 3}
\theta'_{4C}-{5\over 3}\theta'_{2L}\;\right]\nonumber\]
\[ J_\triangle =2\pi\left[\;\triangle_{2L}+{5\over 3}\triangle_Y-
{8\over 3}\triangle_{3C}+\triangle'_{2L}+\triangle'_{2R}-
2\triangle'_{4C}\;\right]\nonumber\]
\b K_\triangle =2\pi\left[\;{5\over 3}(\triangle_Y-\triangle_{2L})+
\triangle'_{2L}+{2\over 3}\triangle'_{4C}-{5\over 3}\triangle'_{2L}
\;\right]\e 
The first, second, and the third terms in the r.h.s. of (2.8)--(2.9) 
represent the one--loop, the multiloop, and the threshold effects, 
respectively.
Each of these contain contributions originating from lower 
scales $\mu=M_Z-M_I$, and higher scales $\mu=M_I-M_U$. We now examine 
the contributions to $\ln{M_I\over M_Z}$ term by term. In the 
presence of the $G_{224P}$ gauge symmetry for 
$\mu\geq M_I,\,\alpha_{2L}(\mu)=\alpha_{2R}(\mu)$. 
Then eq.(2.2) gives\[ a'_{2L}=a'_{2R}\nonumber\]
\[\theta'_{2L}=\theta'_{2R}\nonumber\]\b\triangle'_{2L}
=\triangle'_{2R}\e where the $G_{224P}$ 
symmetry implies
\[\triangle^U_{2L}=\triangle^U_{2R}\nonumber\]
\b\triangle_{2L}^{NRO}=\triangle_{2R}^{NRO}\;,\;\triangle_{2L}^{str}
=\triangle_{2R}^{str}\e 
The restoration of left--right discrete symmetry in the presence of 
$SU(4)_C$ in $G_{224P}$ plays a crucial role in giving rise to 
vanishing contribution due to every type of higher scale corrections.

\subsection{One--loop contributions}

Using (2.12) we find that $B_U$ and $A_U$ are proportional to 
each other,
\b B_U={2\over 3}(a'_{2L}-a'_{4C})=-{1\over 3}A_U\e
\[ D={5\over 3}(a_{2L}-a_Y)A_U-\left(\;{8\over 3}a_{3C}-a_{2L}-{5
\over 3}a_Y\;\right)B_U\nonumber\]
\b ={4A_U\over 9}(3a_{2L}+2a_{3C}-5a_Y)\e 
Then $B_U$ or $A_U$ cancel out from the denominator and the numerator 
of the one-loop term in (2.9) leading to
\[\left(\;\ln{M_I\over M_Z}\;\right)_{one\;loop}={12\pi\over\alpha d}
\left(\;\sin^2\theta_W-{1\over 2}+{1\over 3}{\alpha\over\alpha_S}
\;\right),\nonumber\]
\b d=3a_{2L}+2a_{3C}-5a_Y\e 
The fact that $a'_i\,(i=2L,2R,4C)$ are 
absent from (2.16) demonstrates that the scale $M_I$ is independent 
of the one--loop contribution to the gauge couplings  
emerging from higher scales, $\mu =M_I-M_U$. But these 
coefficients do not cancel out from $\ln {M_U\over M_Z}$, which 
assumes the form,
\b\ln{M_U\over M_Z}={12\pi\over\alpha d}\left(\;\sin^2\theta_W
-{1\over 2}+{\alpha\over 3\alpha_S}\;\right)   
+X\e
\widetext
\FL\[ X={6\pi\over\alpha d}\left[\;a_{3C}
\left(\;1-{8\over 3}\sin^2\theta_W\;\right)+\right.\nonumber\]
\FR\b\left.a_{2L}\left(\;{5\over 3}{\alpha\over\alpha_S}
-1+\sin^2\theta_W\;\right)+{5\over 3}a_Y
\left(\;\sin^2\theta_W-{\alpha\over\alpha_S}\;\right)\;\right]\Big/(a'_
{4C}-a'_{2L})\e
\narrowtext 
The first term in the r.h.s. of (2.17) is the one--loop contribution 
in (2.16).
\par 
We also note that for any standard weak doublet $(H)$
\[ a_{3C}^{(H)}=0\;,\;3a_{2L}^{(H)}=5a_Y^{(H)}\nonumber\] 
which keeps the one--loop term in 
(2.16) unchanged. Thus, the scale $M_I$ is predominantly unaffected 
by the presence of any number of light doublets with masses 
$<M_I$, degenerate or nondegenerate.

\subsection{Two--loop and higher--loop effects:}

Using the second term in the r.h.s. of (2.9), 
(2.14) and (2.15), the coefficients $a'_i$ and terms 
containing $\theta'_i$ cancel out, leading to
\[\left(\;\ln{M_I\over M_Z}\;\right)_{multiloop}=
{K_\theta A_U-J_\theta B_U\over D}\nonumber\]
\b={2\pi\over d}(5\theta_Y-3\theta_{2L}-2\theta_{3c})\e 
showing that all multiloop contributions to the gauge couplings 
originating from $\mu=M_I-M_U$ are absent in $\ln{M_I\over M_Z}$. 
But these multiloop effects do not cancel out from the unification mass,
\b\left(\;\ln{M_U\over M_Z}\;\right)_{multiloop}=\left(\;\ln
{M_I\over M_Z}\;\right)_{multiloop}+X_\theta\e 
where the first term in the r.h.s. of (2.20) is the same as in (2.19),
\widetext
\FL\[ X_\theta={9\pi\over 4d(a'_{4C}-a'_{2L})}\times\left[
\;\left\{\;{5\over 3}\left(\;\theta_{2L}+{5\over 3}\theta_Y-{8\over 
3}\theta_{3C}\;\right)+{10\over 3}(\theta'_{2L}-\theta'_{4C})\;\right\}
(a_{2L}-a_Y)\right.\nonumber\]
\FR\b\left.-\left(\;{8\over 3}a_{3C}-a_{2L}-{5\over 3}a_Y\;\right)
\left\{\;{5\over 3}(\theta_Y-\theta_{2L})+{2\over 3}
(\theta'_{4C}-\theta'_{2L})\;\right
\}\;\right]\e
\narrowtext

\subsection{Threshold effects}

Including threshold effects at $\mu=M_Z,\,M_I$ and $M_U$, 
we separate $J_\triangle$ and $K_\triangle$ into three different parts
\[ J_\triangle=J_\triangle^ U+J_\triangle^I+J_\triangle^Z\nonumber\]
\[ K_\triangle=K_\triangle^U+K_\triangle^I+K_\triangle^Z\nonumber\] 
where
\[ J_\triangle^U=2\pi\left(\;\triangle^U_{2L}+\triangle^U_{2R}
-2\triangle^U_{4C}\;\right)\;,\nonumber\]
\[ J_\triangle^i=2\pi\left(\;\triangle^i_{2L}+{5\over 3}\triangle^i_Y
-{8\over 3}\triangle^i_{3C}\;\right)\;,\;i=I,Z,\nonumber\]
\[ K_\triangle^U=2\pi\left(\;\triangle^U_{2R}+{2\over 3}
\triangle^U_{4C}-{5\over 3}\triangle^U_{2L}\;\right)\;,\nonumber\]
\b K^i_\triangle={10\pi\over 3}\left(\;\triangle^i_Y-\triangle^i_{2L}
\;\right)\;,\;i=I,Z\e 
Using the parity restoration constraint gives
\[ K_\triangle^U={4\pi\over 3}\left(\;\triangle^U_{4C}
-\triangle^U_{2L}\;\right)=-{1\over 3}J_\triangle^U\nonumber\] 
and
\b J_\triangle^UB_U-K_\triangle^UA_U=0\e 
Using (2.23) in the third term in (2.9) gives
\b\left(\;\ln{M_I\over M_Z}\;\right)_{threshold}=-{9\over 4d}
\left(\;K^I_\triangle+{J^I_\triangle\over 3}+K^Z_\triangle
+{J^Z_\triangle\over 3}\;\right)\e 
Thus, it is clear that the would be dominant source of 
uncertainty due to GUT--threshold effects has 
vanished from $\ln{M_I\over M_Z}$ which contains 
contributions from only lower thresholds at $\mu=M_Z$ and $\mu=M_I$. 
But the GUT--threshold contributions do not cancel out from 
$\ln{M_U\over M_Z}$ which has the form
\b\left(\;\ln{M_U\over M_Z}\;\right)_
{threshold}=\left(\;\ln{M_I\over M_Z}\;\right)_{threshold}
+X_\triangle\e 
where
\FL\[X_\triangle=2\pi{\left(\;\triangle^U_{4C}-\triangle^U_{2L}
\;\right)\over\left(\;a'_{4C}-a'_{2L}\;\right)}\nonumber\]
\[+{9\over 4d}\left[\;\left(\;K^I_\triangle+K^Z_
\triangle\;\right)\left(\;{8\over 3}a_{3C}-a_{2L}-{5\over 3}a_Y
\;\right)\right.\nonumber\]
\FR\b\left.-{5\over 3}\left(\;J^I_\triangle+J^Z_\triangle\;\right)
\left(\;a_{2L}-a_Y\;\right)\;\right]\Big/
\left(\;a'_{4C}-a'_{2L}\;\right)\e

\subsection{Gravitational smearing and string threshold effects}

In the presence of left--right discrete symmetry in 
$G_{224P},\,\triangle^{NRO}_{2L}=\triangle^{NRO}_{2R}$ and 
$\triangle^{str}_{2L}=\triangle^{str}_{2R}$. The analysis of Sec.(C) 
holds true in these cases also leading to
\[ J^{NRO}_\triangle B_U-K^{NRO}_\triangle A_U=0\nonumber\]
\[ J^{str}_\triangle B_UY-K^{str}_\triangle A_U=0\nonumber\]
\[\left(\;\ln{M_I\over M_Z}\;\right)_p=0\;,\;p=NRO,string\nonumber\]
\[\left(\;\ln{M_U\over M_Z}\;\right)_{NRO}={2\pi
\over\alpha_G}{\left(\;\in_{2L}-\in_{4C}\;\right)\over
\left(\;a'_{4C}-a'_{2L}\;\right)}\nonumber\]
\b\left(\;\ln{M_U\over M_Z}\;\right)_{str}=2\pi
{\left(\;\triangle^{str}_{4C}-\triangle^{str}_{2L}\;\right)
\over\left(\;a'_{4C}-a'_{2L}\;\right)}\e 
Thus, the theorem is proved demonstrating explicitly that 
$\ln{M_I\over M_Z}$ does not have any modification due to corrections 
to the gauge coupling constants at higher scales for $\mu >M_I$. 
When the Higgs scalars, fermions or gauge bosons of the full $G_{224P}$
representations are taken into account, their contributions to 
$\ln{M_I\over M_Z}$ vanish exactly. The origin behind all cancellations 
is the $G_{224P}$ symmetry and the relation between the gauge couplings,
\[{1\over\alpha_Y(\mu)}={3\over 5}{1\over\alpha
_{2L}(\mu)}+{2\over 5}{1\over\alpha_{4C}(\mu)}\;,\;\mu\geq M_I\nonumber\] 
Since no specific particle content has been used in proving the 
vanishing corrections, the theorem holds true without or with SUSY 
and also in superstring based models.
\par 
Another stability criterion on $M_I$ with respect to 
contributions from lower scale corrections is that, up to one--loop 
level, it remains unchanged by the presence of any number of light 
weak doublets having masses from $M_Z$ to $M_I$.
\par 
The other byproduct of this analysis is on the stability of $M_U$ 
with respect to $16_H+\overline{16}_H$ pairs. In all correction terms 
for $\ln(M_U/M_Z)$ , the higher scale one--loop coefficients appear 
in the combination $a'_{4C}-a'_{2L}$. We note that for any $16_H$ (or 
$\overline{16}_H$)
\[ (\;a'_{4C}\;)_{16_H}=(\;a'_{2L}\;)_{16_H}\] 
which keeps the value of $a'_{4C}-a'_{2L}$ unaltered. Thus, the 
value of $M_U$ is almost unaffected by the presence of any 
number of pairs of $16_H\oplus\overline{16}_H$ between $\mu=M_I-M_
{GUT}$. This has relevance for SUSY $SO(10)$ and string inspired models.

\section{PREDICTIONS IN NONSUSY $SO(10)$}

The stability of $M_I$ in nonSUSY $SO(10)$, under the 
variation of $\eta^{(1)}$ in (2.5) was demonstrated in Ref.[21] by 
accurate numerical 
estimation. According to the present theorem $\ln{M_I\over M_Z}$ is 
not only independent of the 5--dimentional operator and $\eta^{(1)}$, 
but also of other higher dimentional operators in (2.5) 
and parameters arising from the GUT scale. Similarly, the 
vanishing GUT threshold correction to $M_I$, obtained in the 
accurate numerical evaluation of Ref.[22], is a part of the 
present theorem. Imposing the parity restoration criteria for 
$\mu\geq M_I$ [23], the minimal nonSUSY $SO(10)$ with \underline{54},   
\underline{126} and\underline{10} representations, 
$\sin^2\theta_W=0.2316\pm 0.0003,\,\alpha_s(M_Z)=0.118\pm 0.0007$, 
and $\alpha^{-1}(M_Z)=127.9\pm 0.1$ predicts [21--23],
\[M_I=10^{13.6\pm 0.16^{+0.5}_{-0.4}}\;{\rm GeV}\;,\nonumber\]
\[M_U=10^{15.02\pm 
0.25\pm 0.48\pm 0.11(0.25)}\;{\rm GeV}\nonumber\] 
Where the first (second) uncertainties are due to those in the 
input parameters (threshold effects). In the case of $M_I$, the 
threshold uncertainties are due to those at $M_Z$ and $M_I$ 
thresholds only. The third uncertainty due to 5--dimensional 
operator in (2.5), which is absent in $M_I$, has been calculated for 
$\eta^{(1)}=\pm 5(\pm 10)$. Inspite of addition of a number of 
extra \underline{126} and \underline{10} dimensional Higgs fields 
to build a model 
for degenerate and seesaw contributions to the neutrino masses 
in $SO(10)$ introducing $SU(2)_H$ horizontal symmetry, the scale 
$M_I$, according to the present theorem, is identical to that 
in the minimal model with the same predictions on the 
nondegenerate neutrino masses [24]. The proton lifetime predictions 
in the minimal model including NRO contribution is
\[\tau_{p\rightarrow e^+\pi^0}=1.44\times 10^{32.1\pm 0.7\pm 1.0
\pm 1.9\pm 0.45(1.0)}\;{\rm yrs}.\nonumber\] 
which might be testified by the 
next generation of experiments.

\section{INTERMEDIATE SCALE IN SUSY $SO(10)$}

In the conventional SUSY $SO(10)$ employing the Higgs supermultiplets
\underline{54}, $16_H\oplus\overline{16}_H$ and \underline{10}, 
in the usual fashion, it is impossible to 
achive $M_I$ substantially lower than $M_U$. When 
$126_H\oplus\overline{126}_H$ are used instead of 
$16_H\oplus\overline{16}_H$, no intermediate gauge group 
containing $SU(4)_C$ has been found to be possible in Ref.[25]. 
But the possibilities of other 
intermediate gauge symmetries in string inspired SUSY $SO(10)$ 
including $G_{2213}\,(g_{2L}\neq g_{2R})$ have been demonstrated 
[25,26] by using extra light $G_{2213}$--submultiplets not needed 
for spontaneous symmetry breaking, but predicted to be 
existing in the spectrum [19].
\par 
In the present analysis, in addition to the usual \underline{54} 
with all components at the GUT scale, the pair $16_H+\overline{16}_H$ 
with desired components at $G_{224P}$ breaking scale, and the 
bidoublet $\phi(2,2,1)\subset\underline{10}$ near $M_Z$ 
while (2,2,6) is at $M_U$, we examine the effects of other components 
in \underline{45}, or in $16_H+\overline{16}_H$ not absorbed by 
intermediate scale gauge bosons, being lighter and having masses 
between 1TeV --$M_I$.
\par 
The adjoint representation \underline{45} contains the left--handed 
triplet $\sigma_L(3,1,1)$, the right--handed triplet $\sigma_R(1,3,1)$ 
and also $\sigma^{(C)}(1,1,15)$ under $G_{224P}$. Under the standard 
gauge group, $\sigma_R$ and $\sigma^{(C)}$
decompose as 
\[\sigma_R(1,3,1)=\sigma^{(+)}_R(1,1,1)+\sigma^{(-)}_R(1,-1,1)
+\sigma^{(0)}_R(1,0,1)\nonumber\]
\FL\[\sigma^{(C)}(1,1,15)=\sigma^{(C)}_3\left(\;1,{2\over 3},3\;\right)
+\sigma^{(C)}_{\overline{3}}\left(\;1,-{2\over 3},\overline{3}\;\right)+
\nonumber\]
\FR\[\sigma^{(C)}_8(1,0,8)+\sigma^{(C)}_S(1,0,1)\nonumber\] 
The representation $16_H$ contains the $G_{224P}$ submultiplets 
$\chi^{(L)}(2,1,4)$ and $\chi^{(R)}(1,2,\overline{4})$ and the 
latter decomposes under SM gauge group as
\FL\[\chi^{(R)}(1,2,\overline{4})=\chi_1^{(R)}(1,-1,1)+\chi_S^{(R)}
(1,0,1)+\nonumber\]
\FR\[\chi^{(R)}_{\overline{3}}\left(\;1,-{2\over 3},\overline{3}
\;\right)+\chi^{(R)'}_{\overline{3}}\left(\;1,-{1\over 3},\overline{3}
\;\right)\nonumber\] 
To make the model simpler, we assume some of these lighter components 
from \underline{45} or the pair $16_H\oplus\overline{16}_H$ to be 
either at $M_C\simeq 1$ TeV while others are at $M_I$. In that case 
all the equations for $\ln{M_I\over M_Z}$ and $\ln{M_U\over M_Z}$    
derived in Sec.II hold with the replacements:
\[\ln{M_I\over M_Z}\rightarrow\ln{M_I\over M_C}\;,\;\ln{M_U\over M_Z}
\rightarrow\ln{M_U\over M_C}\;,\;\theta_i\rightarrow\theta^C_i
\nonumber\]
\[a_i\rightarrow a_i^c\;(i=Y,2L,3C)\;{\rm and}\;d\rightarrow d_C
\nonumber\] 
in (2.15)--(2.16). In addition, there are contributions to the mass scales 
due to evolutions  from $M_Z-M_C$. We present them here 
only upto one--loop. The two--loop, threshold, and gravitational 
corrections will be estimated elsewhere [27]
\FL\[\left(\;\ln{M_I\over M_C}\;\right)_{one-loop}=
\nonumber\]
\FR\[{12\pi\over\alpha d_C}\left(\;\sin^2\theta_W-{1\over 2}
+{\alpha\over 3\alpha_s}\;\right)-R\;\ln{M_C\over M_Z}\nonumber\]
\b\left(\;\ln{M_U\over M_C}\;\right)_{one-loop}=
\left(\;\ln{M_I\over M_C}
\;\right)_{one-loop}+X_C+Y\e where
\widetext
\FL\[Y={5\over 8d_C(a'_{4C}-a'_{2L})}\nonumber\]
\FR\[\left[\;\left(\;a^C_{2L}-a^C_Y
\;\right)\left(\;3a_{2L}+5a'_Y-8a_{3C}\;\right)-(a_{2L}-a_Y)
\left(\;3a^C_{2L}+5a'^C_Y-8a^C_{3C}\;\right)\;\right]
\times\ln{M_C\over M_Z}\nonumber\]
\narrowtext
\[d_C=d(a_i\rightarrow a^c_i)=3a_{2L}^c+2a_{3C}^c-5a_Y^c\nonumber\]
\[X_C=X(a_i\rightarrow a_i^c)\nonumber\]
\[R={d\over d_c}\nonumber\] 
We find that when the components under the standard gauge group given 
in Table I are at $M_C\simeq 1$ TeV, 
the intermediate mass scale $M_I=5\times 10^{12}-2\times 10^{14}$ GeV 
can be achieved with $M_U=M_{str}\simeq 6\times 10^{17}$ GeV. 
It has been emphasized that the $SU(3)_C$--octet and $SU(2)_L$--weak 
triplet being in the standard model adjoint 
representation and continuous moduli of strings, have a natural 
justification to keep them light [17]. In our case 
$\sigma^\pm,\,\sigma_3,\;\sigma_{\overline{3}}$ and 
$\sigma^{(c)}$ belong to the adjoint representations (1,3,1) and
(1,1,15) of
$G_{224}$ which in turn are contained in the adjoint representation
$\underline{45}\subset SO(10)$. One 
set of our solutions in Table I corresponds to the first three of 
them being as light as $M_C\simeq 1$ TeV while the fourth component, 
the $SU(3)_C$--octet component in $\sigma^c(1,1,15)$ is at $M_I$. 
We have also found a completely different type of 
solution where the $SU(2)_R$--triplet components and 
$\chi_3\oplus\chi_{\overline{3}}\subset 16_H+\overline{16}_H$, 
but not absorbed by $SU(4)_C$ gauge bosons, are near 1 TeV. In that 
case all the components in $\sigma^c(1,1,15)$ are at $M_I$. The 
neutrinos acquire small Majorana masses by seesaw mechanism using 
$SO(10)$ singlets as explained in Ref.[26]. None of the lighter 
scalar degrees of freedom near 1 TeV are needed to acquire vacuum 
expectation values as the spontaneous symmetry breakings of  
gauge symmetries like $SO(10),\,G_{224P}$, and $G_{213}$ occur 
following the standard procedure through the vacuum expectation 
values of well known scalar components which are neutral under 
the residual gauge groups.
\par 
The left--handed neutrinos acquire 
small Majorana neutrino masses via seesaw mechanism where the 
right--handed neutrino mass $M_N$, rather then $M_I$, occurs in the 
seesaw formula, in both SUSY [26] and nonSUSY theories. But since 
$M_N$ is of the same order as $M_I$ with $M_N\leq M_I$ in a
large class of models, the right--handed Majorana mass is also 
made correspondingly uncertain whenever $M_I$ is affected by 
larger uncertainties, especially due to the GUT--threshold effects 
with nondegenerate components of scalar representations [8] and
gravitational effects due to higher dimensional operators [12,21]. 
This occurs in models where parity is broken at the GUT scale, but 
$G_{224}$ or $G_{2213}$ with $g_{2L}\neq g_{2R}$ [8], or even 
$SU(2)_L\times U(1)_R\times SU(4)_C\,(\equiv G_{214})$ [29], 
breaks at the intermediate scale. With $G_{2213P}$ at the 
intermediate scale, these corrections do not vanish, although 
they are reduced. But in the $SO(10)$ and other GUTs, or string 
inspired models with $G_{224P}$ (but not $G_{2213P}$) surviving down to 
the intermediate scale, all major sources of uncertainties 
emerging from higher--scale corrections are absent in $M_I$ and, 
therefore, correspondingly in $M_N$, even though the latter is 
still undetermined within one order of magnitude below $M_I$. It is 
to be emphasized that in such models, the order--of--magnitude 
estimation of right--handed Majorana neutrino masses are much more 
accurate as compared to other models with intermediate scales. 
Consequently, the left--handed--Majorana--neutrino--mass prediction 
is more precise in these models. Further it is not true that 
imposition of the left--right symmetry at the intermediate 
scale always leads to vanishing higher--scale corrections. 
The vanishing correction occurs only in the presence of the 
left--right symmetric $G_{224P}$--gauge symmetry for $\mu >M_I$. 
Mohapatra [30] has proved a theorem on vanishing corrections due to
GUT--threshold effects originating from degenerate components of 
$SO(10)$--Higgs representations in the presence  of other type of 
gauge symmetry. The present theorem emphasizes vanishing corrections 
due to all sources emerging from $\mu >M_I$ in the presence of 
$G_{224P}$ only.

\section{SUMMARY AND CONCLUSIONS}

We have shown that all higher scale corrections on the 
intermediate--scale prediction $(M_I)$, corresponding to the 
$G_{224P}$ gauge symmetry breaking, vanish exactly. Such corrections 
are due to one--loop, two--loop and higher loop effects, 
GUT--threshold and gravitational smearing effects originating from 
higher--dimentional operators. In string inspired SUSY GUTs, 
the string--loop threshold effects have also vanishing contributions 
on $M_I$. In nonSUSY $SO(10)$ models, the intermediate scale has been 
predicted earlier and we emphasize that $M_I\simeq 10^{13.6}$ GeV 
is quite stable leading to more precise neutrino mass predictions. 
The predicted proton lifetime can be testified by future experiments. 
The $G_{224P}$ symmetry having only two gauge 
couplings guarantees unification, but the problem in SUSY $SO(10)$ 
is the realization of $M_I\ll M_U$. We find solutions to this 
problem with $M_I\simeq 5\times 10^{12}-2\times 10^{14}$ GeV 
and $M_U\simeq M_{str}\simeq 6\times 10^{17}$ GeV for small 
$\alpha_S(M_Z)$ provided certain states in the adjoint representation 
\underline{45} and/or $16_H+\overline{16}_H$ have masses near 1 TeV. 
The light states in $16_H+\overline{16}_H$ may emerge naturally 
from the modes not absorbed by heavy $SU(2)_R\times SU(4)_C$ 
gauge bosons. String--scale unification might be possible in case of 
another intermediate symmetry, such as $G_{2213}$, with parity broken 
at the GUT scale, when the submultiplet $\sigma^c(1,1,0,8)$ is at 
the intermediate scale [28]; but only in 
the present case of $G_{224P}$ intermediate symmetry, the scale $M_I$ 
has all higher--scale corrections vanishing and neutrino mass 
predictions in SUSY $SO(10)$ are expected to be more precise.

\acknowledgements 

The author thanks J.C.Pati, R.N.Mohapatra, 
G.Senjanovic and A.Yu.M.Smirnov for useful discussions. 
Financial support and hospitality from the Theory Group, International 
Center for Theoretical Physics, Trieste, are gratefully acknowledged. 
The author also acknowledges the grant of a research 
project SP/S2/K-09/91 from the Department of Science and Technology, 
New Delhi.

\mediumtext

\begin{table}
\caption{Predictions for mass scales in string inspired $SO(10)$ model}
\begin{tabular}{ccccccc} 
SM submultiplets&SM submultiplets&$G_{224P}$ 
submultiplets&$a^c_i$&$a'_i$&$M_I$&$M_U$\\ 
$M_Z-M_I$&$M_C-M_I$&$M_I-M_U$&&&
(GeV)&(GeV)\\ 
\tableline \\
{$\phi_u,\phi_d$}&
{$\begin{array}{ccc}
\sigma_R^\pm,&\sigma_3,&\sigma_{\overline{3}}\\
&{\rm or}&\\
\sigma_R^\pm,&\chi_3,&\chi_{\overline{3}}
\end{array}$}&
{$\begin{array}{ccc}
\sigma_L,&\sigma_R,&\sigma^c,\\
\chi_L,&\chi_R,&\bar{\chi}_L,\\
&\bar{\chi}_R,\phi&
\end{array}$}&
{$\left(\begin{array}{c}
{47\over 5}\\ 1\\ -2
\end{array}\right)$}&
{$\left(\begin{array}{c}7\\ 2\end{array}\right)$}&
$10^{12.5}$&$10^{17.6}$\\ \\ \hline  \\
{$\phi_u,\phi_d$}&
{$\begin{array}{ccc}
\chi_1,&\chi_3,&\chi_{\bar{3}},\\
&\chi'_3&\\
\end{array}$}&
{$\begin{array}{ccc}
\chi_L,&\chi_R,&\bar{\chi}_L,\\
&\bar{\chi}_R,\phi&\\
\end{array}$}
&{$\left(\;\begin{array}{c}
{42\over 5}\\ 1\\ 
-{3\over 2}\end{array}\;\right)$}
&{$\left(\;\begin{array}{c}5\\ -2\\
\end{array}\;\right)$}&$10^{14.3}$&$10^{17.8}$\\ \\
\end{tabular}
\end{table}

\end{document}